\documentclass[aps,pra,preprint,groupedaddress,showpacs,%
tightenlines,%
nofootinbib,floatfix]{revtex4-1}
\usepackage{graphicx}
\usepackage{amsmath,amssymb,amsfonts,bm}
\usepackage{array,dcolumn}
\usepackage{hyperref}

\renewcommand{\vec}[1]{\bm{\mathrm{#1}}} % 3D-vectors

\begin{document}
%======================================================================

\title{Muonic-atom scattering from hydrogenic liquids in an incoherent
  approach}

\author{Andrzej Adamczak}
\email{andrzej.adamczak@ifj.edu.pl}
\affiliation{Institute of Nuclear Physics Polish Academy of Sciences,
  Radzikowskiego 152, PL-31342~Krak\'ow, Poland}
\author{Andrzej Z.~G\'orski} 
\email{andrzej.gorski@ifj.edu.pl}
\affiliation{Institute of Nuclear Physics Polish Academy of Sciences,
  Radzikowskiego 152, PL-31342~Krak\'ow, Poland}

\date{\today}

\begin{abstract}
  The differential cross sections for low-energy muonic hydrogen atom
  scattering in liquid hydrogenic targets have been calculated in the
  incoherent approximation using the Van Hove response function.
  A~simple model of liquids and the available experimental parameters
  have been employed for a~description of the diffusive and vibrational
  modes in these targets. At collision energies below about 10~meV, the
  obtained cross sections are very different from the analogous cross
  sections for scattering in hydrogenic gases.
\end{abstract}


\maketitle

%---------------------
\section{Introduction}
\label{sec:intro}
%---------------------

Many experiments in low-energy muon physics were performed in the liquid
mixtures of hydrogen isotopes. In particular, such hydrogenic targets
were employed for studying the muon-catalyzed $pd$ and $pt$ fusion in
the muonic molecules $pd\mu$ and $pt\mu$ (see, e.g.,
\cite{doed63,bles63,peti90,baum93,mark93}). The yields of different
products from these fusion reactions strongly depend on the populations
of hyperfine states of the muonic molecules. These yields are functions
of the concentrations of deuterium and tritium admixtures in the H$_2$
targets, which is known as the Wolfenstein-Gershtein
effect~\cite{wolf60,zeld60}. This is due to a complicated chain of
processes from the Coulomb capture of the negative muons to formation of
the muonic molecules. The time evolution of the spin and kinetic energy
of the $d\mu$ and $t\mu$ atoms is of particular importance.  In order to
compare the theoretical fusion rates with the experimental data, it is
necessary to accurately describe the kinetics of muon-catalyzed fusion.
Detailed kinetics calculations~\cite{peti90,baum93,mark93}, were
performed using then available cross sections for scattering of the
muonic atoms from the hydrogen-isotope
nuclei~\cite{buba87,brac89,chic92}. However, in the considered
experiments, these processes take place in the molecular liquid targets.
One may thus expect that the realistic cross sections are very different
from that for the scattering from free nuclei or molecules, which has
already been shown in the case of solid hydrogenic
targets~\cite{adam07}. An estimation of the differential cross sections
for muonic atom scattering from the liquid hydrogenic targets in a
simple incoherent approximation is the aim of this work. The obtained
results can be applied for an accurate simulation of the new
experimental study of muon-catalyzed $pt$ fusion, which is underway at
JINR in Dubna.  This experiment has been planned in order to clarify the
large discrepancies between theory and experiment in the $pt\mu$ case,
and to observe more fusion channels (e.g.,
$pt\mu\to{}^{4}\mathrm{He}+e^{+}+e^{-}$) for the first
time~\cite{bogd12,bogd13}.

A method of estimating the differential cross sections for muonic
hydrogen atom scattering in liquid hydrogenic targets in the incoherent
approximation is presented in Sec.~\ref{sec:approx}. Some examples of
the calculated cross sections for the liquid hydrogen, deuterium, and
hydrogen with a small admixture of deuterium or tritium, at the
temperature $T=22$~K and saturated-vapor pressure, are shown in
Sec.~\ref{sec:results}. A summary of the obtained results and
conclusions are given in Sec.~\ref{sec:concl}.

%---------------------------------------------------------------------
\section{Method of calculation}
\label{sec:approx}
%---------------------------------------------------------------------

Since a muonic hydrogen atom $a\mu$ ($a=p$, $d$, or $t$) is a small
neutral object, the cross section for $a\mu$ scattering in a condensed
target can be calculated using the methods that were developed for a
description of the neutron scattering. In the incoherent approximation,
the partial differential cross section per a~single target molecule can
be expressed in terms of the incoherent fraction
$\mathcal{S}_i(\vec{\kappa},\omega)$ of the Van Hove response function
$\mathcal{S}(\vec{\kappa},\omega)$~\cite{vanh54}
\begin{equation}
  \label{eq:xsec_part}
  \frac{\partial^2\sigma}{\partial\Omega\partial\varepsilon'} =
  \frac{k'}{k}\,\sigma_\text{mol}\,\mathcal{S}_i(\vec{\kappa},\omega)\,,
\end{equation}%
The squared amplitude for $a\mu$ scattering from a free hydrogenic
molecule is denoted here by~$\sigma_\text{mol}$, the energy transfer
$\omega$ and momentum transfer $\vec{\kappa}$ to the condensed target
are defined as
\begin{equation}
  \label{eq:e_m_trans}
  \omega=\varepsilon-\varepsilon'-\Delta{}E, \qquad 
  \vec{\kappa}=\vec{k}-\vec{k}'\,,
\end{equation}
where $\varepsilon$ and $\varepsilon'$ are the initial and final kinetic
energies of the atom; vectors $\vec{k}$ and $\vec{k}'$ denote the
corresponding momenta. The energy transfer due to internal transitions
in the atom and the target molecule is denoted by $\Delta{}E$.
Equation~(\ref{eq:xsec_part}) is exact for the incoherent processes of
internal transitions in the atom (spin flip) or molecule
(rotational-vibrational excitations). In the case of strictly elastic
scattering, the approximation $\mathcal{S}\approx\mathcal{S}_i$ is valid
for higher momentum transfers, when the coherent processes are
relatively small. It has been shown in Ref.~\cite{adam07} that the total
cross sections for $a\mu$ scattering in solid hydrogens are practically
unaffected by the coherent processes at collision energy
$\varepsilon\gtrsim{}1$--2~meV. In Eq.~(\ref{eq:xsec_part}), the
rotational-vibrational transitions in the target molecules and the
spin-flip transitions in the scattered atom $a\mu$ are taken into
account in the free-molecule squared amplitudes $\sigma_\text{mol}$,
which were calculated in Ref.~\cite{adam06} for all combinations of the
hydrogen isotopes. These calculations employed the amplitudes for $a\mu$
scattering from the bare hydrogen-isotope nuclei~\cite{brac89,chic92} as
the input.

The response function $\mathcal{S}_i$ in Eq.~(\ref{eq:xsec_part})
describes the translative motion of molecules in a liquid target. The
studies of liquid H$_2$ and D$_2$ using slow neutron scattering showed
that both the diffusive and collective modes are present in the dynamics
of these quantum liquids (see, e.g.,
Refs.~\cite{berm93,berm99,berm00,mukh98} and references therein). These
studies revealed the phonon spectra similar to those characteristic for
polycrystalline powders. Thus, despite the lack of a periodic structure
in the liquid hydrogens, it is reasonable to describe their properties
in terms of the Debye temperature for a certain range of the momentum
transfers. At lowest energies, the scattering in liquids is dominated by
the diffusive motion of target particles, which results in the presence
of a broad quasielastic peak centered at the incident energy. At the
liquid hydrogen density, the interactions between the neighboring
molecules are important. This leads to a recoil-less scattering at
lowest $\kappa$ and $\omega$. For example, such effect was observed in
the lowest rotational excitation of H$_2$ in collision with
neutrons~\cite{momp97}.

Many advanced theoretical models and computer programs were developed
for a description of neutron scattering from liquid hydrogen and
deuterium (see, e.g., Refs.~\cite{gran04,guar15,guar16}).  These models
lead to different forms of the response function, which is usually
convoluted with a response from the integral degrees of freedom of the
molecules H$_2$ and D$_2$. Such convolution involves the spin
correlations that are characteristic for the neutron scattering from
these molecules and different from the analogous case of muonic hydrogen
atom scattering. In our approach, the spin correlations and
rotational-vibrational structure of the molecules are already taken into
account in the square amplitudes~$\sigma_\text{mol}$. For these reasons,
below we evaluate the response function for the specific case of muonic
atom scattering, using a simple general model from Ref.~\cite{love84}.

In liquid hydrogens, the diffusive motion at small~$\kappa$ is well
described~\cite{berm93} by the Langevin equation
\begin{equation}
  \label{eq:Langevin}
  M \frac{d^2\vec{R}}{dt^2} = -\zeta M \frac{d\vec{R}}{dt}
  +\vec{F}_s(t) \,,
\end{equation}
where $M$ is the molecular mass, $\vec{R}$ is a position of the
molecule, $\zeta$ denotes the strength of a frictional force due to the
movement in liquid, and $\vec{F}_s$ is a stochastic force connected with
the scattering from other molecules. The coefficient $\zeta$ is related
to the self-diffusion coefficient $D_s$ by the Einstein relation
\begin{equation}
  \label{eq:Einstein}
  \zeta = k_B T /(M D_S) \,,
\end{equation}
in which $k_B$ denotes Boltzmann's constant.  Equations
(\ref{eq:Langevin}) and (\ref{eq:Einstein}) lead to the following
diffusion contribution~\cite{love84}:
\begin{equation}
  \label{eq:S_diff}
  \mathcal{S}_\text{diff}(\kappa,\omega) = 
  \frac{1}{\pi}\exp(-2W) \beta_T\omega \, [n_B(\omega)+1] \,
  \frac{D_s\kappa^2}{\omega^2+(D_s\kappa^2)^2}
\end{equation}
to the response function $\mathcal{S}_i$. Function $\exp(-2W)$ denotes
the Debye-Waller factor and the Bose factor $n_B(\omega)$ is defined as
\begin{equation}
  \label{eq:Bose_fac}
  n_B = [\exp(\beta_T\omega)-1]^{-1} ,
\end{equation}
where $\beta_T=1/(k_BT)$. Equation~(\ref{eq:S_diff}) can be applied for
small $\kappa$ and $\omega$, apart from the limit $\kappa$,
$\omega\to{}0$. The width $\Delta_\text{diff}$ at the half maximum of
the quasi-elastic Lorentzian factor in Eq.~(\ref{eq:S_diff}) equals
$2D_s\kappa^2$. Since the neutron experiments revealed that
$\Delta_\text{diff}$ is almost constant at higher momentum transfers, we
fix its value at $\kappa>\kappa_\text{max}$.  For example,
$\kappa_\text{max}\approx{1.7}$~\AA$^{-1}$ in the case of 20-K liquid
deuterium~\cite{berm93}.

The collective-motion contribution $\mathcal{S}_\text{pho}$ to
$\mathcal{S}_i$ can be described using the following incoherent phonon
expansion for a harmonic crystal~\cite{vanh54,love84}:
\begin{equation}
  \label{eq:S_vib}
  \mathcal{S}_\text{vib}(\kappa,\omega) = \exp(-2W) \sum_{n=1}^{\infty}
  g_n(\omega) (2W)^n / n! \,,
\end{equation}
without the strictly elastic term~$\delta(\omega)$, which is
characteristic for scattering in solids. Functions $g_n$, which describe
the subsequent $n$-phonon processes, and the exponent $2W(\kappa^2)$ of
the Debye-Waller factor are given in Ref.~\cite{adam07} for the
isotropic Debye model of a solid. We estimate the effective Debye
temperature $\Theta_D$ using the relation:
\begin{equation}
  \label{eq:T_Deb}
  \Theta_D=\frac{h c_s}{k_B}\left(\frac{3}{4\pi}\frac{N_A}{V}\right)^2 ,
\end{equation}%
where $h$ is Planck's constant, $c_s$ denotes the sound velocity, $N_A$
is Avogadro's constant, and $V$ represents the molar volume. The
corresponding Debye energy is defined as $\omega_D=k_B\Theta_D$.

\begin{figure}[htb]
  \centering
  \includegraphics[width=8cm]{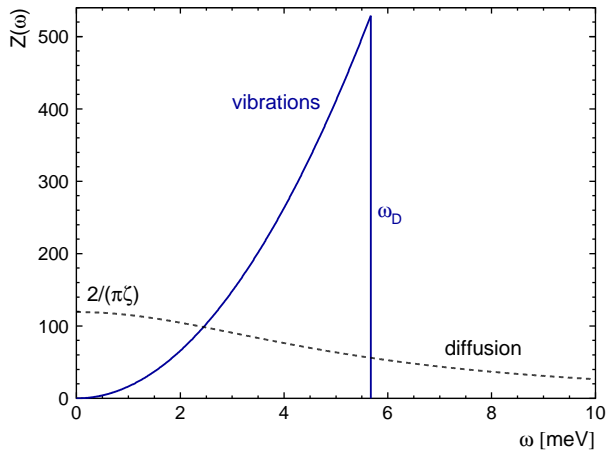}
  \caption{Generalized frequency spectrum $Z(\omega)$ (in arbitrary
    units) for the simple model of 22-K liquid hydrogen.}
  \label{fig:z_H2}
\end{figure}

In theory of liquids, a generalized frequency spectrum $Z(\omega)$ was
introduced~\cite{egel62}, which is analogous to the density of
vibrational states in solids. The Langevin equation~(\ref{eq:Langevin})
leads to the following contribution from the diffusive motion:
\begin{equation}
  \label{eq:Z_diff}
  Z_\text{diff}(\omega) = \frac{2}{\pi\zeta} 
  \frac{\zeta^2}{\omega^2+\zeta^2} 
\end{equation}%
to $Z(\omega)$. A contribution from the collective vibrations is given
by the standard Debye density $Z_\text{vib}(\omega)$ of vibrational
states. These two contributions are shown in Fig.~\ref{fig:z_H2} for the
22-K liquid
hydrogen at saturated-vapor pressure. The values of parameters $D_s$,
$c_s$, and $V$ are taken from Ref.~\cite{soue86}. The presented plot
exhibits characteristic features, which are observed in liquids.  The
spectrum at lowest $\omega$ is determined by the diffusive modes. In
particular, $Z(\omega)$ tends to a finite value when $\omega$ approaches
zero. For solids, this limit equals zero. The Debye frequency $\Theta_D$
corresponds to the large peaks in liquids, which are due to the
finite-frequency collective oscillations. The tail at large~$\omega$,
which is apparent in the experiments, is consistent the Enskog theory
for a fluid of hard spheres~\cite{resi77}.

\begin{figure}[htb]
  \centering
  \includegraphics[width=8cm]{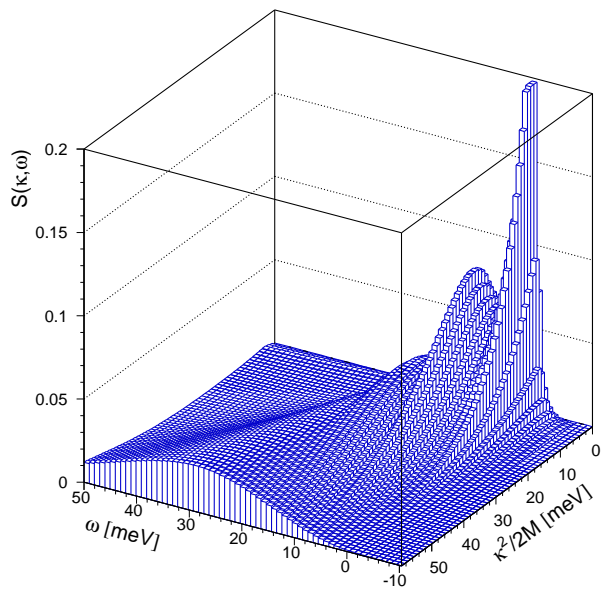}
  \caption{Incoherent response function~(\ref{eq:S_i}) for the 22-K
    liquid hydrogen at saturated-vapor pressure.}
  \label{fig:resp_H2}
\end{figure}

For the first estimation of the cross sections for muonic atom
scattering in liquid hydrogens, we use a sum of the functions
(\ref{eq:S_diff}) and (\ref{eq:S_vib})
\begin{equation}
  \label{eq:S_i}
  \mathcal{S}_i(\kappa,\omega) \approx 
  \mathcal{S}_\text{diff}(\kappa,\omega) 
  +\mathcal{S}_\text{vib}(\kappa,\omega) 
\end{equation}
as a fair approximation of the total response function.
Figure~\ref{fig:resp_H2} presents the function~(\ref{eq:S_i})
for the 22-K liquid H$_2$ at saturated-vapor pressure. In this plot,
one can clearly distinguish the diffusive contribution at the lowest
$\omega$, and then the subsequent contributions from the one-phonon and
two-phonon processes.

For large $\kappa$ and $\omega$, any system (apart from liquid helium at
lowest temperatures) is accurately described by the asymptotic Gaussian
form~\cite{vanh54,love84}
\begin{equation}
  \label{eq:S_asym}
  \mathcal{S}_i(\kappa,\omega)=\exp[-(\omega-\omega_R)^2/\Delta_R^2]\,,
  \quad \Delta_R = \sqrt{8\mathcal{E}_T\omega_R/3}
\end{equation}
in which $\omega_R=\kappa^2/(2M)$ denotes the recoil energy and
$\mathcal{E}_T$ is the mean kinetic energy of a hydrogenic molecule in
the considered liquid. In solid and liquid hydrogens, this energy is
significantly greater than the Maxwellian energy $\tfrac{3}{2}k_BT$
owing to large zero-point vibrations of the light hydrogenic molecules
in these quantum systems~\cite{zopp01}. The mean kinetics energy can be
estimated using the following formula
\begin{equation}
  \label{eq:E_T}
  \mathcal{E}_T = \tfrac{3}{2}\int_0^\infty d\omega \, 
  Z_\text{vib}(\omega)\, \omega \left[n_B(\omega)+\tfrac{1}{2}\right]\,.
\end{equation}
The asymptotic response function~(\ref{eq:S_asym}) leads to a correct
free-molecule limit at large energy transfers.

%------------------------------------------------------------
\section{Results of calculations}
\label{sec:results}
%------------------------------------------------------------

Our calculations of the cross sections have been performed for liquid
hydrogen and deuterium at 22~K and saturated-vapor pressures, which
corresponds to the conditions of the experiments performed at
PSI~\cite{peti90,baum93} and JINR~\cite{bogd12}. At 22~K, practically
all symmetric molecules H$_2$ and D$_2$ are in the ground rotational
state $K=0$. Thus, we are dealing with the liquid parahydrogen
(para-H$_2$) and orthodeuterium (ortho-D$_2$).  Only freshly prepared
targets are statistical mixtures of the rotational states $K=0$ and
$K=1$, due to a slow rotational deexcitation of the symmetric
molecules~\cite{soue86}.  Such targets are often called normal hydrogens
and labeled as nH$_2$ and nD$_2$, respectively.

\begin{figure}[htb]
  \centering
  \includegraphics[width=8cm]{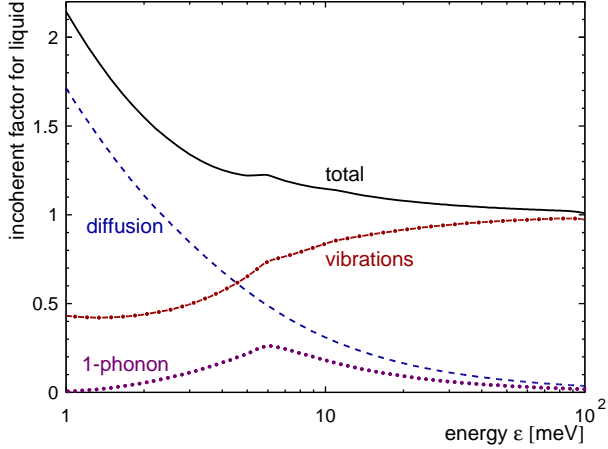}
  \caption{Function $\mathcal{C}_\text{inc}$ for $p\mu$ scattering in
    the 22-K liquid hydrogen. The contributions to
    $\mathcal{C}_\text{inc}$ from the diffusive and vibration modes are
    shown separately. The label ``1-phonon'' denotes the process of
    one-phonon creation.}
  \label{fig:zfactor_H2}
\end{figure}

An estimation of the Debye temperature using Eq.~(\ref{eq:T_Deb}) and
the parameters $c_s$ and $V$ from Ref.~\cite{soue86} results in
$\Theta_D\approx{}66$~K and $\omega_D\approx{}5.7$~meV, for both
hydrogen and deuterium. For our conditions, the value of self-diffusion
coefficient~$D_s$ equals $1.12\times{}10^{-4}$~cm$^2$/s for hydrogen and
$0.49\times{}10^{-4}$~cm$^2$/s for deuterium~\cite{soue86}. These
parameters have been applied for plotting Figs.~\ref{fig:z_H2}
and~\ref{fig:resp_H2}, and for calculating the cross sections. In order
to illustrate contributions from the diffusive and vibration modes to
the total cross section, the following function
\begin{equation}
  \label{eq:C_inc}
  \mathcal{C}_\text{inc}(\varepsilon) = \frac{1}{4\pi} 
  \left(\frac{M_{a\mu}}{\mathcal{M}}\right)^2 \int d\Omega\, 
  d\varepsilon'\, \frac{k'}{k} \mathcal{S}_i(\kappa,\omega)
\end{equation}
is shown in Fig.~\ref{fig:zfactor_H2}. In Eq.~(\ref{eq:C_inc}),
$M_{a\mu}$ denotes the mass of $a\mu$ and $\mathcal{M}$ is a reduced
mass of the system which consists of the impinging muonic atom and a
single hydrogenic molecule.
Function $\mathcal{C}_\text{inc}$ is a ratio of the cross section for
$a\mu$ scattering from this molecule in the liquid target and the
corresponding cross section for the same free molecule. In
definition~(\ref{eq:C_inc}), it is assumed that the latter cross section
is constant. Function $\mathcal{C}_\text{inc}$ falls from the maximal
value of $(M_{a\mu}/\mathcal{M})^2$, at $\varepsilon\to{}0$, to the
free-molecule limit of~1 at $\varepsilon\gtrsim{}100$~meV. The
quasielastic diffusive mode is most important at lowest energies. The
vibration mode includes annihilation and creation of one phonon and many
phonons. The plotted one-phonon contribution to $\mathcal{C}_\text{inc}$
has a maximum at the Debye energy.

\begin{figure}[htb]
  \centering
  \includegraphics[width=8cm]{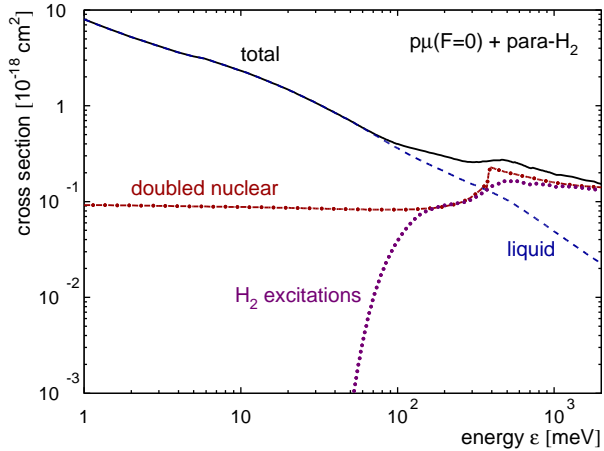}
  \caption{Total cross section for $p\mu(F=0)$ scattering in the 22-K
    liquid para-H$_2$. The label ``liquid'' denotes scattering without
    internal excitations of the target molecules. The label ``H$_2$
    excitations'' represents a~contribution from the
    rotational-vibrational excitations of the H$_2$ molecule. The label
    ``doubled nuclear'' denotes the doubled cross section for the
    elastic scattering $p\mu(F=0)+p$.}
  \label{fig:xppp11_liq_contrib}
\end{figure}

The total cross section for scattering of the $p\mu$ atom from liquid
parahydrogen is shown in Fig.~\ref{fig:xppp11_liq_contrib}. 
The atom is in the ground spin state $F=0$ during the scattering
process. The contributions to the cross section from the molecular
motion in para-H$_2$ and from the rotational-vibrational excitations of
the target molecule H$_2$ are shown separately. The latter contribution
begin to appear at $\varepsilon\approx{}40$~meV, which corresponds to
the first rotational excitation $K=0\to{}2$ of~H$_2$. The transition
$K=0\to{}1$ is forbidden for the state $F=0$ of $p\mu$. The doubled
total cross section for $p\mu(F=0)$ scattering from a bare
proton~\cite{brac89} is shown for a comparison. In this Figure, three
regimes of scattering are visible. For $\varepsilon\lesssim{}100$~meV,
effects of the target molecule interactions with the neighboring H$_2$
molecules are important. Above this energy, the scattering passes to the
free-molecule regime. Finally, above about 1~eV (depending on the choice
of hydrogen isotopes~\cite{adam06}), the free-nuclei regime is achieved.

\begin{figure}[htb]
  \centering
  \includegraphics[width=8cm]{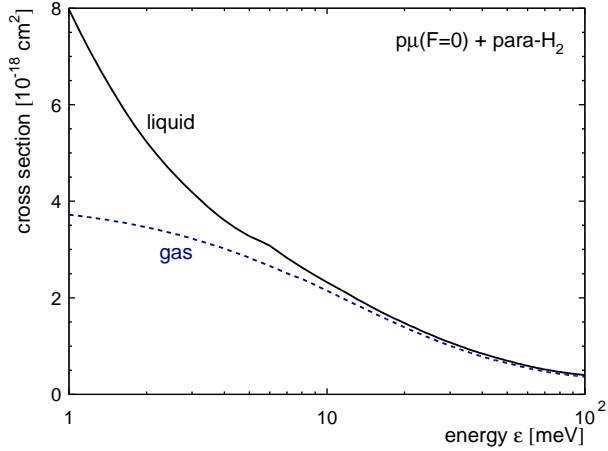}
  \caption{Total cross sections for $p\mu(F=0)$ scattering in the 22-K
    liquid and gaseous para-H$_2$.}
  \label{fig:xppp11_liq_k0}
\end{figure}%
The total cross sections for $p\mu(F=0)$ scattering in the liquid and
gaseous parahydrogen are shown in Fig.~\ref{fig:xppp11_liq_k0}. One can
see that the cross sections begin to significantly diverge below 10~meV.
The differential cross sections for $a\mu$ scattering from hydrogenic
targets are strongly anisotropic. This can be conveniently shown using
the transport cross sections
\begin{equation}
  \label{eq:sig_trans}
  \sigma_\text{tr}(\varepsilon) = \int d\Omega \, d\varepsilon' \,
  (1-\cos\vartheta) 
  \frac{\partial^2\sigma}{\partial\Omega\partial\varepsilon'} \,,
\end{equation}
where $\vartheta$ denotes the scattering angle. These cross sections are
employed for a simple description of the slowing down process of
different particles.
\begin{figure}[htb]
  \centering
  \includegraphics[width=8cm]{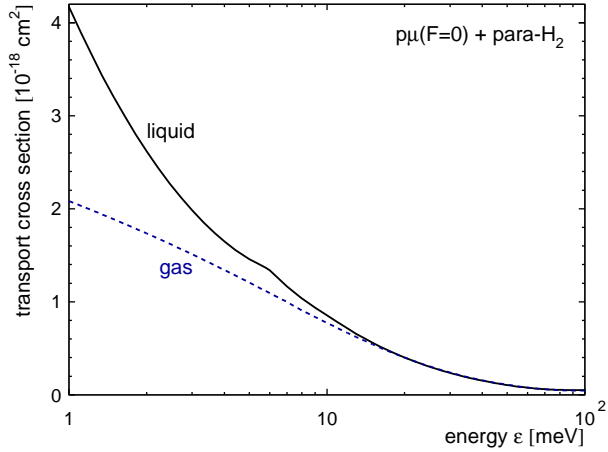}
  \caption{Transport cross sections for $p\mu(F=0)$ scattering in the
    22-K liquid and gaseous para-H$_2$.}
  \label{fig:xtr_ppp11_liq_k0}
\end{figure}%
The transport cross sections for $p\mu(F=0)$ scattering in the liquid
and gaseous para-H$_2$ are plotted in Fig.~\ref{fig:xtr_ppp11_liq_k0}.

\begin{figure}[htb]
  \centering
  \includegraphics[width=8cm]{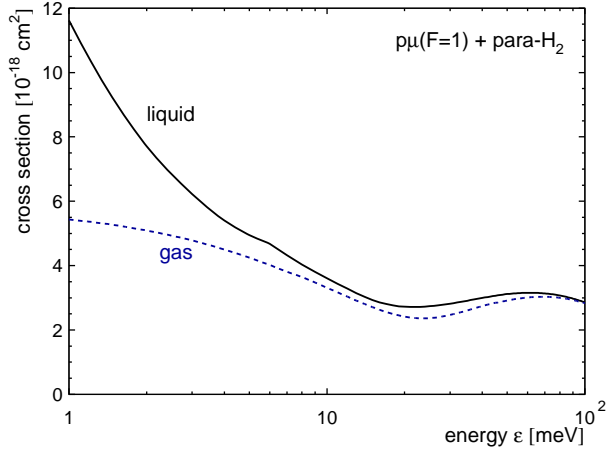}
  \caption{The same as in Fig.~\ref{fig:xppp11_liq_k0} for $p\mu(F=1)$.}
  \label{fig:xppp22_liq_k0}
\end{figure}

\begin{figure}[htb]
  \centering
  \includegraphics[width=8cm]{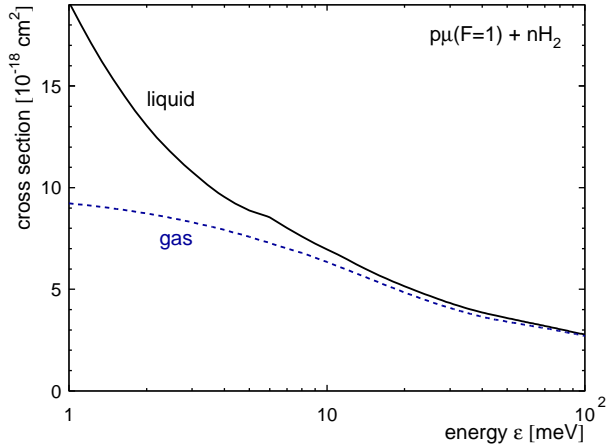}
  \caption{Total cross sections for $p\mu(F=1)$ scattering in liquid and
    gaseous nH$_2$.}
  \label{fig:xppp22_liq_stat}
\end{figure}

The cross section for scattering of the $p\mu$ atom in the excited spin
state $F=1$ significantly differs from that for the ground spin state
$F=0$. The amplitudes of $p\mu(F=0)$ and $p\mu(F=1)$ scattering from
a~proton have the opposite signs and quite different magnitudes.
Moreover, in the case of $p\mu(F=1)$ scattering from the H$_2$ molecule,
the ortho-para rotational transitions are allowed due to the proton
exchange between the atom and the H$_2$ molecule. A difference of the
both cross sections is apparent while comparing
Fig.~\ref{fig:xppp11_liq_k0} with Fig.~\ref{fig:xppp22_liq_k0}.
Other differences between the cross sections are due to the scattering
in hydrogenic targets with various populations of the molecular
rotational states. The cross sections for the $p\mu(F=1)$ scattering in
liquid and gaseous normal hydrogen are presented in
Fig.~\ref{fig:xppp22_liq_stat} as an example. 
In nH$_2$, a population of the excited rotational state $K=1$ is equal
to 75\%. As a result, the rotational deexcitation $K=1\to{}0$ is
possible, which significantly changes the cross sections at lowest
energies (see Figs.~\ref{fig:xppp22_liq_k0}
and~\ref{fig:xppp22_liq_stat}).

Energies of the hyperfine splitting in the muonic hydrogen atoms
(182~meV for $p\mu$ and 48.5~meV for $d\mu$) are much greater than the
Debye energy of about 6~meV.
\begin{figure}[htb]
  \centering
  \includegraphics[width=8cm]{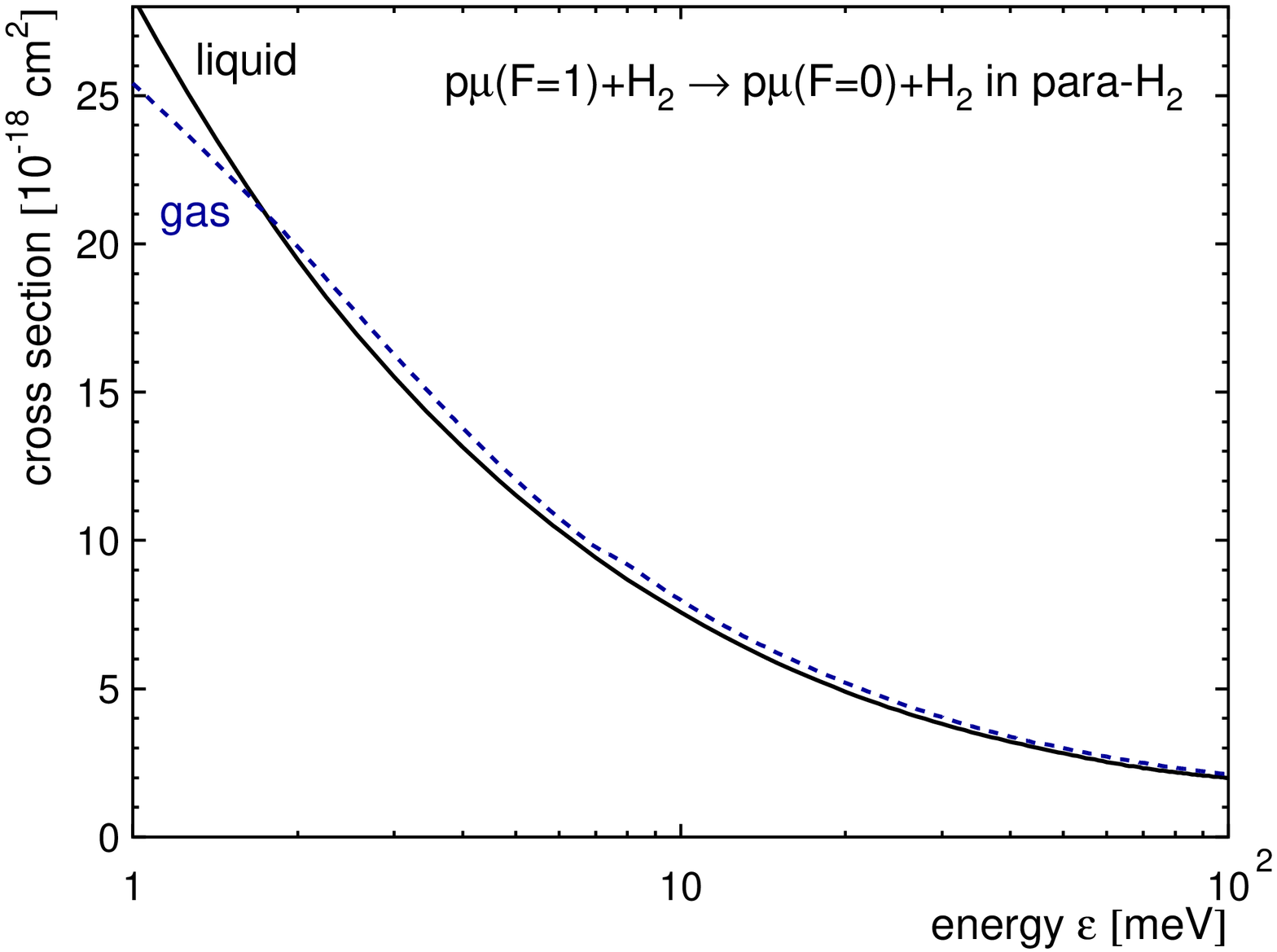}
  \caption{Total cross sections for the spin flip process
    $p\mu(F=1)+\text{H}_2\rightarrow{}p\mu(F=0)+\text{H}_2$ in the 22-K
    liquid and gaseous para-H$_2$.}
  \label{fig:xppp21_liq_k0}
\end{figure}%
Therefore, the energy transfer~$\omega$ in the spin-flip transitions is
relatively large. As a result, the spin-flip cross sections in $a\mu$
scattering from the liquid hydrogens only slightly differ from the
corresponding cross sections for hydrogenic gases. The reaction
$p\mu(F=1)+\mathrm{H}_2\to{}p\mu(F=0)+\mathrm{H}_2$ in the liquid and
gaseous parahydrogen is plotted in Fig.~\ref{fig:xppp21_liq_k0}. An
significant difference between the spin-flip cross sections is apparent
only below 2~meV. Let us note that the effective spin-flip reaction in
the scattering $p\mu+\mathrm{H}_2$ is due to the muon exchange between
the protons~\cite{gers58a}.

\begin{figure}[htb]
  \centering
  \includegraphics[width=8cm]{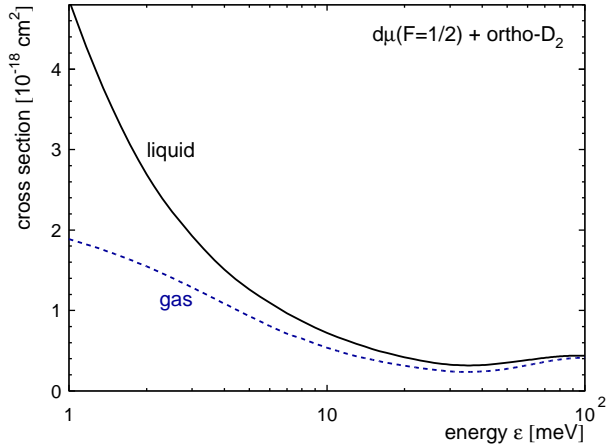}
  \caption{Total cross sections for $d\mu(F=1/2)$ scattering in the 22-K
    liquid and gaseous ortho-D$_2$.}
  \label{fig:xddd11_liq_k0}
\end{figure}

\begin{figure}[htb]
  \centering
  \includegraphics[width=8cm]{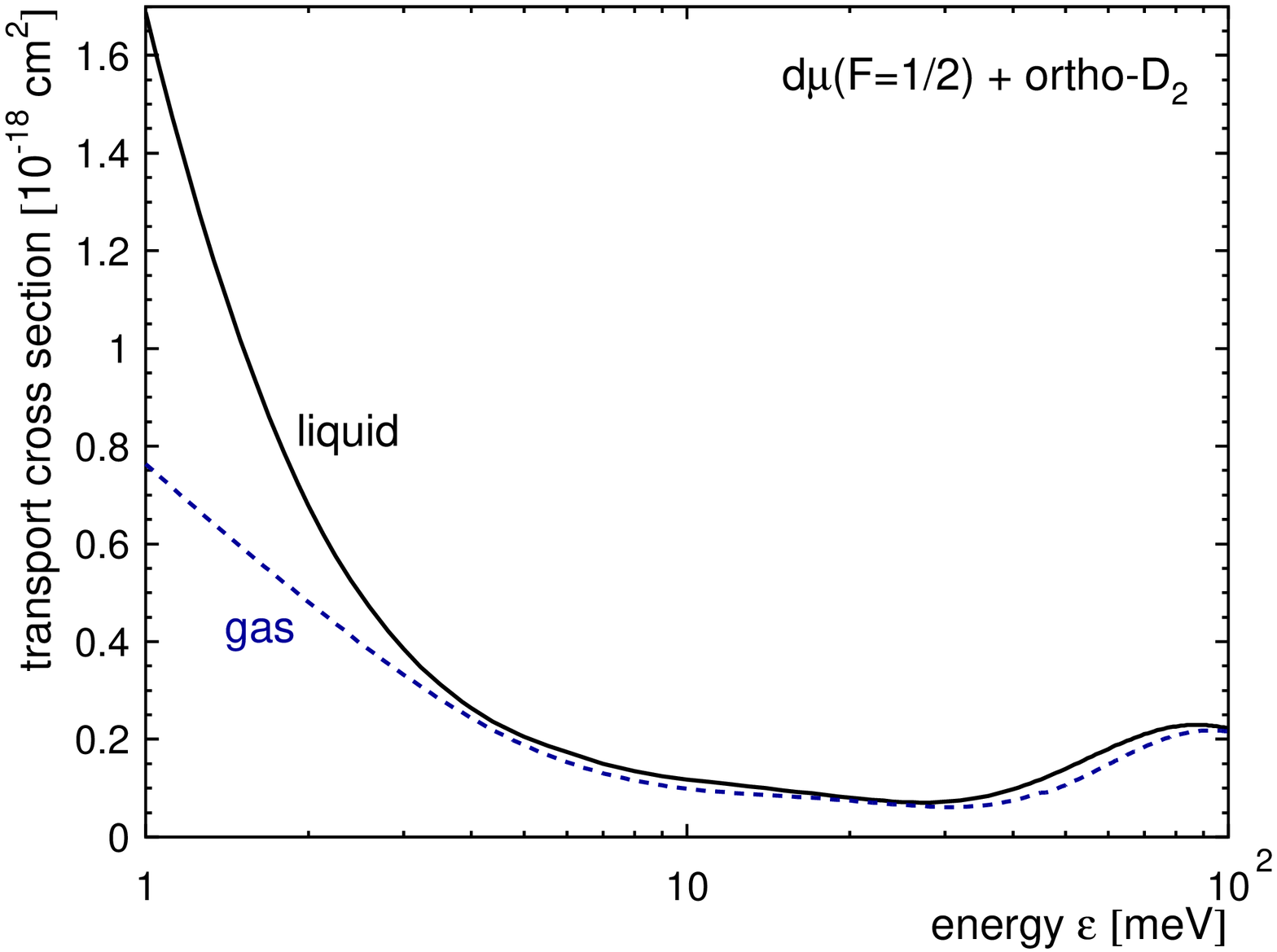}
  \caption{Transport cross sections for $d\mu(F=1/2)$ scattering in the
    22-K liquid and gaseous ortho-D$_2$.}
  \label{fig:xtddd11_liq_k0}
\end{figure}

Figures \ref{fig:xddd11_liq_k0} and \ref{fig:xtddd11_liq_k0} present the
total and transport cross sections for scattering of the $d\mu$ atom in
the ground spin state $F=1/2$ from the liquid and gaseous orthodeuterium
at~22~K.
The corresponding cross sections for $d\mu$ scattering in the upper spin
state $F=3/2$ are similar since the mean amplitudes for the scattering
$d\mu(F=1/2)+d$ and $d\mu(F=3/2)+d$ are quite close. The mean amplitude
denotes here averaging over the total spin of the system $d\mu+d$.
Moreover, the rotational transition $K=1\to{}0$ in the scattering
$d\mu+\mathrm{D}_2$ is allowed for the both states $F=1/2$ and~3/2. As a
result, the cross sections for $d\mu(F=1/2)$ and $d\mu(F=3/2)$
scattering in nD$_2$, where the population of the rotational ortho-state
$K=0$ equals~2/3, are similar.
\begin{figure}[htb]
  \centering
  \includegraphics[width=8cm]{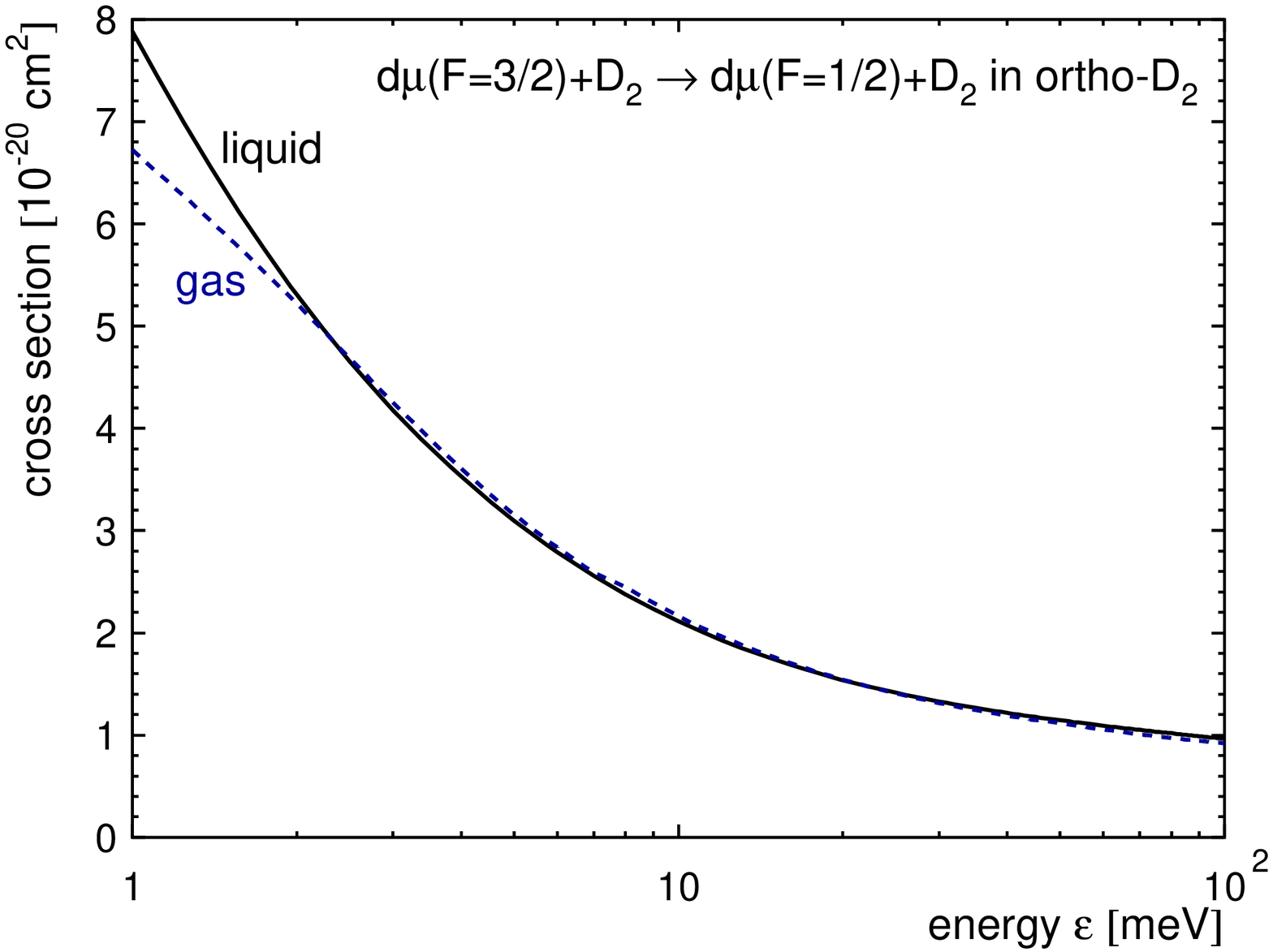}
  \caption{Total cross sections for the spin flip process
    $d\mu(F=3/2)+\text{D}_2\rightarrow{}d\mu(F=1/2)+\text{D}_2$ in the
    22-K liquid and gaseous ortho-D$_2$.}
  \label{fig:xddd21_liq_k0}
\end{figure}%
The cross sections of the spin-flip process
$d\mu(F=3/2)+\text{D}_2\rightarrow{}d\mu(F=1/2)+\text{D}_2$ in the 22-K
liquid and gaseous ortho-D$_2$ are shown in
Fig.~\ref{fig:xddd21_liq_k0}.  A relative difference between these cross
sections is significantly greater (17\% at 1~meV) than in the
$p\mu+\mathrm{H}_2$ case. This is caused by a~smaller hyperfine
splitting in the $d\mu$ atom.

\begin{figure}[htb]
  \centering
  \includegraphics[width=8cm]{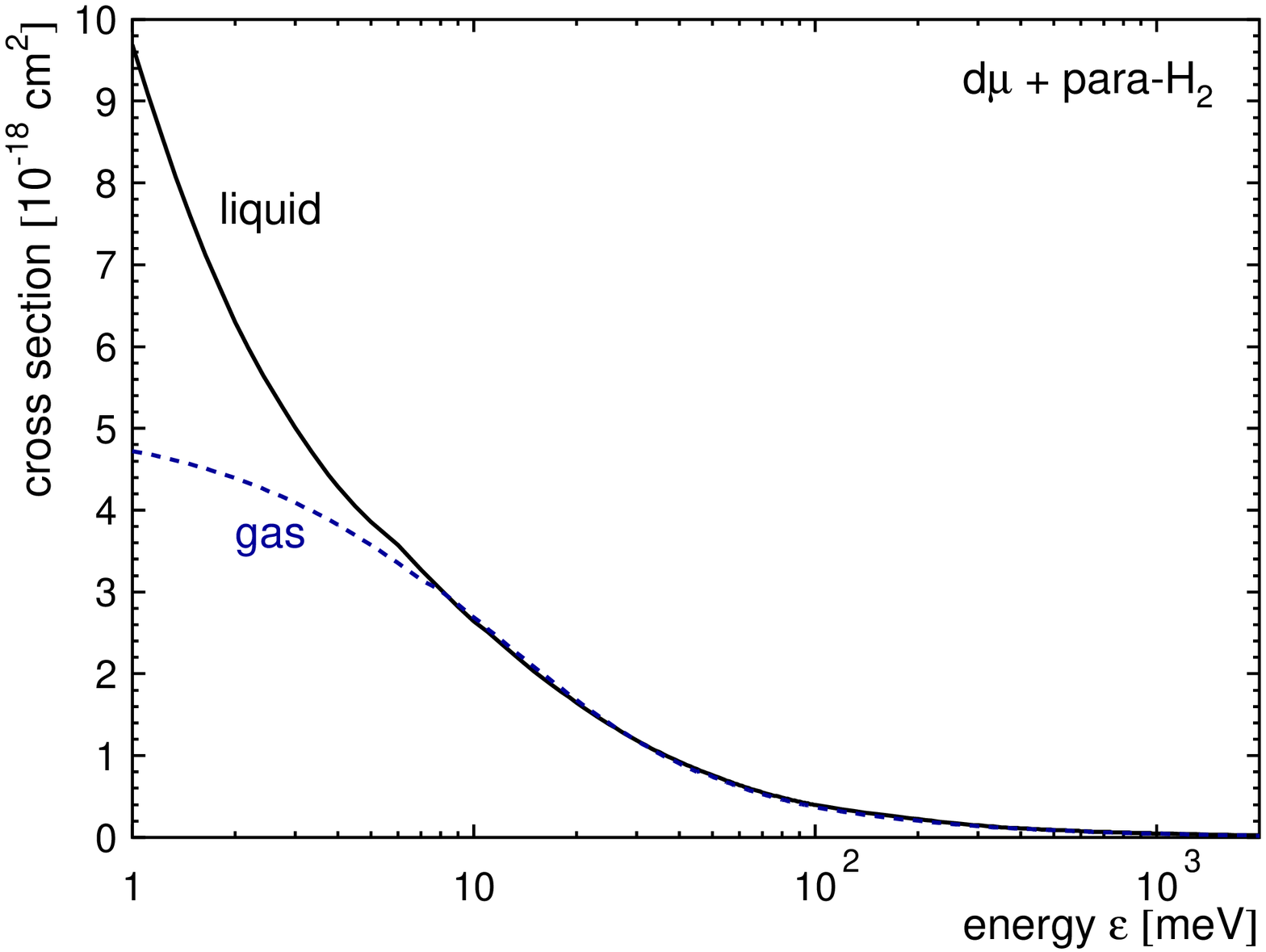}
  \caption{Total cross sections for $d\mu$ scattering from the H$_2$
    molecule in the 22-K liquid and gaseous para-H$_2$.}
  \label{fig:xdpp_liq_k0}
\end{figure}

The experimental studies of fusion reactions in the muonic molecules
$pd\mu$ and $pt\mu$ are often performed in liquid H$_2$ targets with
small admixtures ($\lesssim{}1$\%) of deuterium or
tritium~\cite{bles63,peti90,baum93,bogd12}. Small amounts of the heavier
isotopes do not practically change the density of liquid hydrogen and
the sound velocity. In the case of $d\mu$ scattering in hydrogen with a
small concentration of deuterium, collisions with the H$_2$ molecules are
most frequent. The cross sections for $d\mu$ scattering from the H$_2$
molecules in the 22-K liquid and gaseous H/D mixture are plotted in
Fig.~\ref{fig:xdpp_liq_k0}. 
These cross sections does not depend on the spin of $d\mu$. In such H/D
mixtures, $d\mu$ collision with the deuterons takes place mostly in the
HD molecules, since the concentration of D$_2$ molecules in the
equilibrated H/D mixture is very small.
\begin{figure}[htb]
  \centering
  \includegraphics[width=8cm]{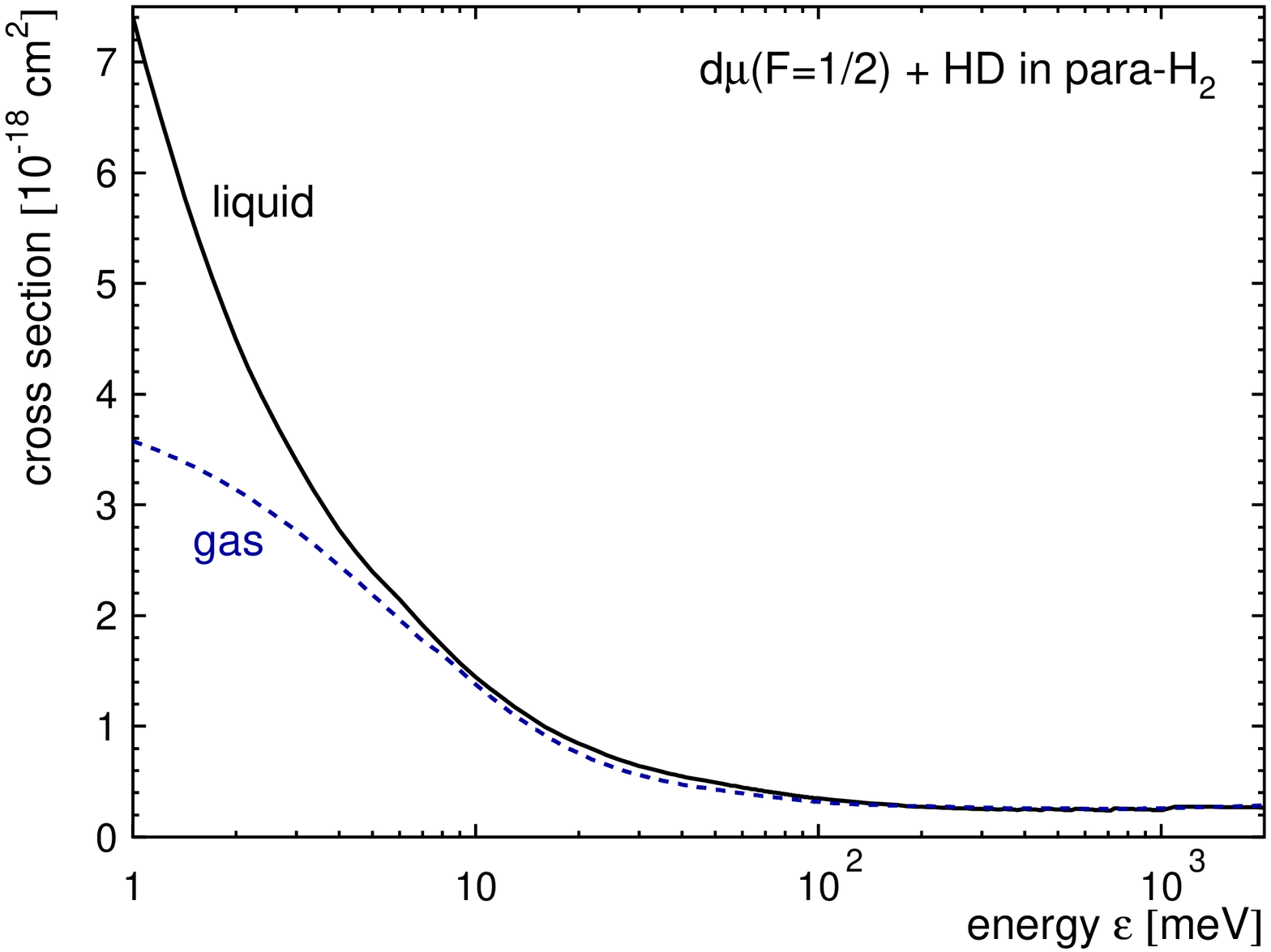}
  \caption{Total cross sections for $d\mu(F=3/2)$ scattering from the HD
    molecule in the 22-K liquid and gaseous para-H$_2$.}
  \label{fig:xddp11_liq_k0}
\end{figure}%
The total cross sections for $d\mu(F=1/2)$ scattering from the HD
molecule in liquid and gaseous hydrogen are plotted in
Fig.~\ref{fig:xddp11_liq_k0}. This molecule is in the ground rotational
state since the asymmetric hydrogenic molecules rapidly deexcite to the
state $K=0$ at low temperatures. For calculating the response function
$\mathcal{S}_\text{diff}$, we have taken
$D_s=0.77\times{}10^{-4}$~cm$^2$/s. This is a value of the diffusion
coefficient for a $D_2$ molecule which diffuses in
liquid~H$_2$~\cite{soue86}. However, this is a fair approximation for
the molecules HD and HT moving in liquid hydrogen, at small
concentrations of deuterium and tritium.
\begin{figure}[htb]
  \centering
  \includegraphics[width=8cm]{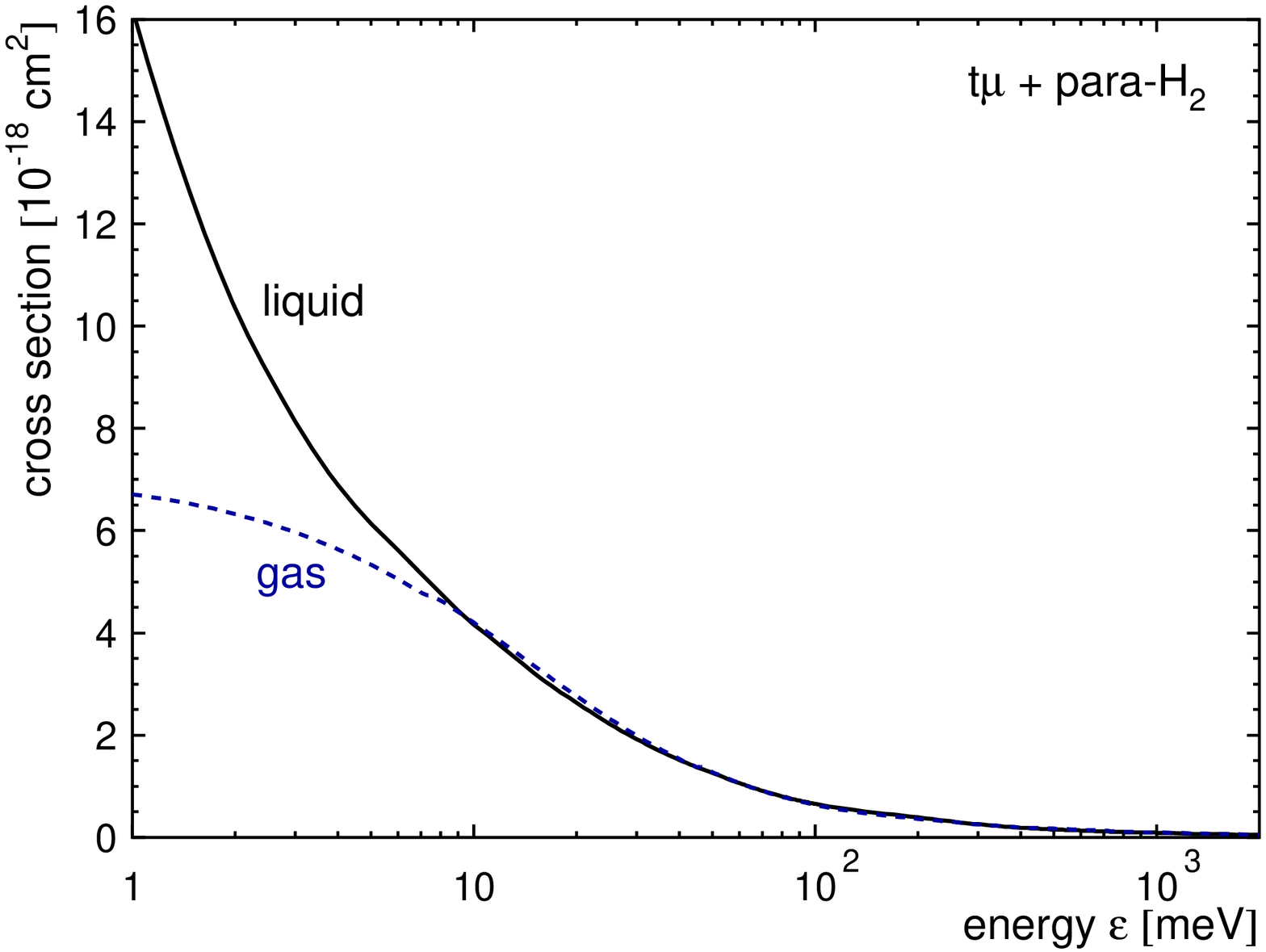}
  \caption{Total cross sections for $t\mu$ scattering from the H$_2$
    molecule in the 22-K liquid and gaseous para-H$_2$.}
  \label{fig:xtpp_liq_k0}
\end{figure}%
\begin{figure}[htb]
  \centering
  \includegraphics[width=8cm]{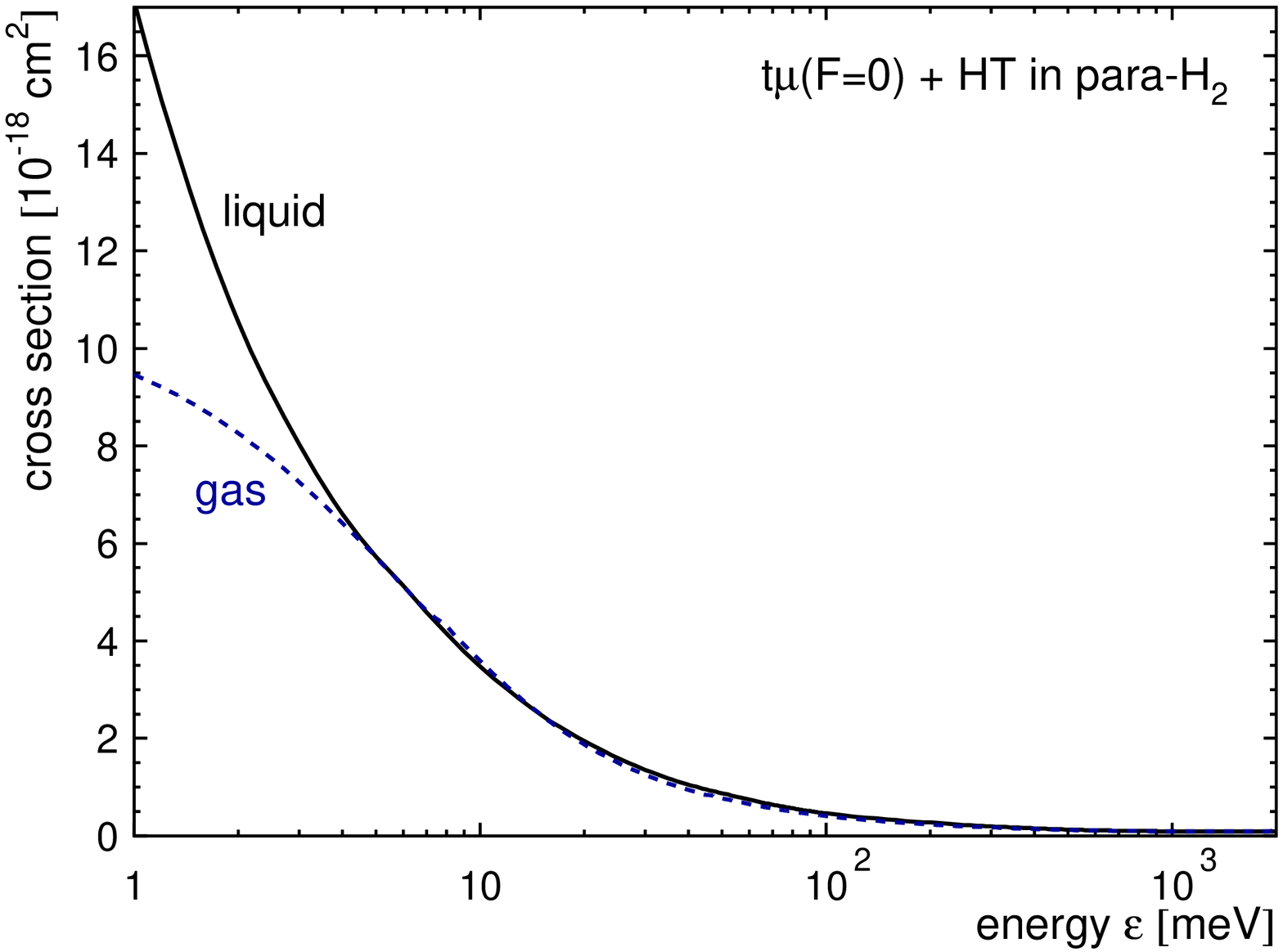}
  \caption{Total cross sections for $t\mu(F=0)$ scattering from the HT
    molecule in the 22-K liquid and gaseous para-H$_2$.}
  \label{fig:xttp11_liq_k0}
\end{figure}%
The total cross sections for $t\mu$ scattering from the H$_2$ and HT
molecules in the liquid and gaseous H/T mixture (small tritium
concentration) are presented in Figs.~\ref{fig:xtpp_liq_k0}
and~\ref{fig:xttp11_liq_k0}.

In the kinetics equations for the muon-catalyzed fusion in H/D/T
mixtures, the rates of different important reactions are employed. In
particular, the spin-flip rates are needed for a correct description of
the time spectra of fusion products.
\begin{figure}[htb]
  \centering
  \includegraphics[width=8cm]{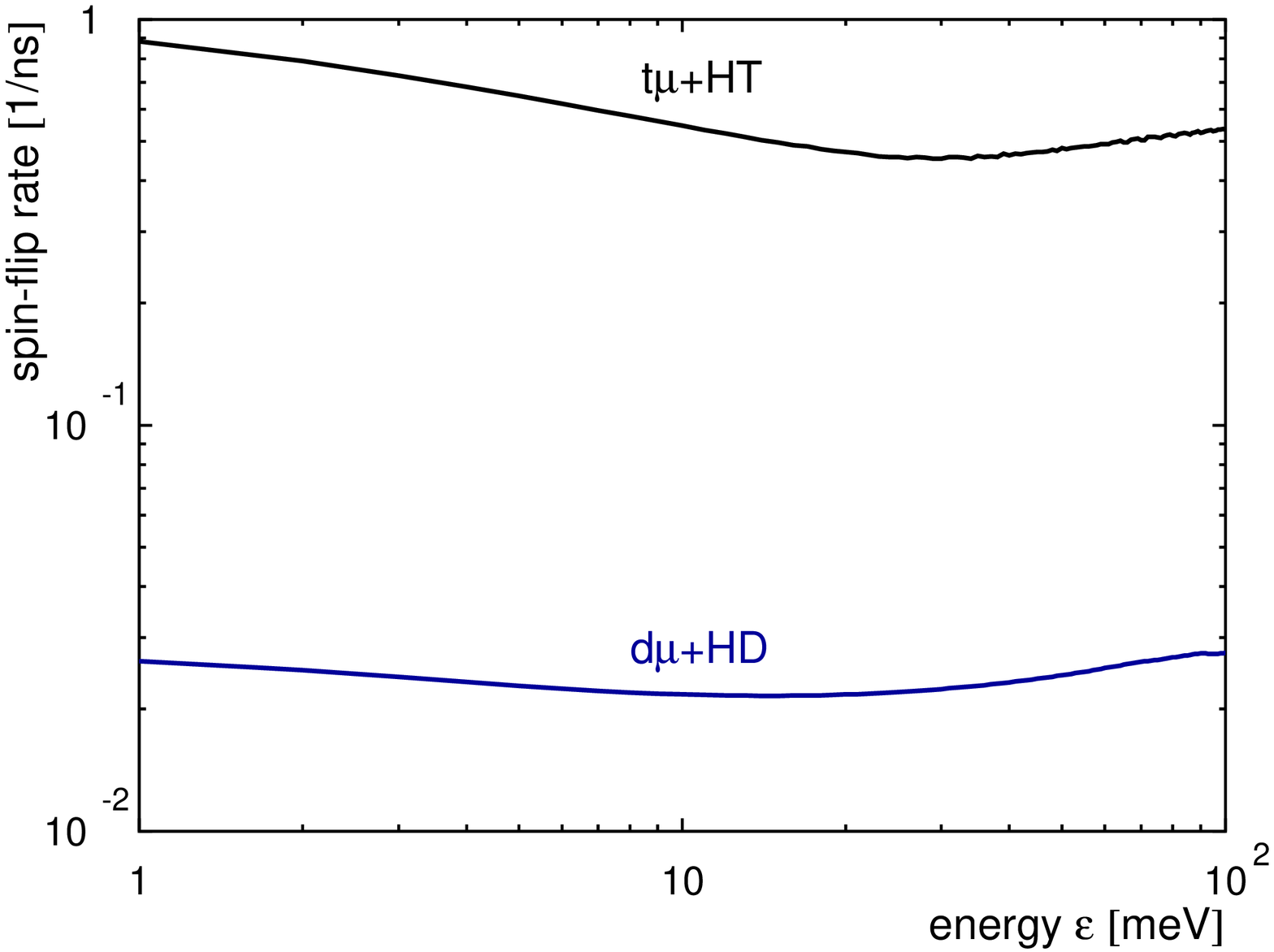}
  \caption{Total rates of spin deexcitation in the scattering
    $d\mu(F=3/2)+$HD and $t\mu(F=1)+$HT in the 22-K liquid hydrogen.}
  \label{fig:rsf21_liq_k0}
\end{figure}%
In Fig.~\ref{fig:rsf21_liq_k0} we present the calculated spin-flip rates
for $d\mu(F=3/2)$ scattering from the HD molecule and $t\mu(F=1)$
scattering from the HT molecule in liquid hydrogen. The rates are
normalized to the density of $2.12\times{}10^{22}$~molecules/cm$^3$,
which is common in low-energy muon physics. This is the number density
of liquid hydrogen at 20.4~K and saturated-vapor pressure.

%---------------------
\section{Conclusions}
\label{sec:concl}
%---------------------

The differential cross sections for low-energy scattering of the muonic
hydrogen atoms from liquid hydrogens have been estimated in the
incoherent approximation. The Van Howe response function and the simple
model of liquid hydrogen, which takes the diffusive and vibrational
degrees of freedom into account, have been employed. Our calculations
show that effects associated with the presence of liquid are very
important for kinetic energies below 10~meV. Above 100~meV, the
corresponding cross sections for scattering in liquid and gaseous
hydrogens are practically identical. In the case of spin-flip reactions,
an appreciable difference between these cross sections is apparent only
below a~few meV, which is caused by a relatively high (compared to the
Debye energy) energy transfers to the target. This difference (12\% for
the scattering $p\mu(F=1)+\mathrm{H}_2$ and 17\% for
$d\mu(F=3/2)+\mathrm{D}_2$ at 1~meV) is however significant for the
spin-dependent processes, before a~steady-state condition is reached.

Thermalization of the energetic ($\sim{}1$~eV) $p\mu$ atoms in a pure
liquid hydrogen is fast ($\lesssim{}10$~ns) owing to high magnitudes of
the scattering cross sections and a large number density of the target
molecules. The same conclusion holds for thermalization of the $d\mu$
atoms in a pure liquid deuterium. As a result, the time spectra, such as
the spectrum of $dd$-fusion products, are affected by properties of the
liquid only at short times ($\lesssim{}10$~ns).

Another situation can be observed in the case of liquid hydrogen with
small admixtures of deuterium and tritium. The negative muons from a
beam are captured mostly by the H$_2$ molecules, which leads to
formation of the $p\mu$ atoms and subsequent formation of the $d\mu$ or
$t\mu$ atoms in the muon exchange reactions $p\mu+d\to{}d\mu+p$ or
$p\mu+t\to{}t\mu+p$. The released energy greatly accelerates the new
atoms. They cannot be effectively slowed down to thermal energies in
scattering from the abundant H$_2$ molecules because of the deep
Ramsauer-Townsend minima in the elastic scattering processes $d\mu+p$
and $t\mu+p$, at energies of a few eV~\cite{chic92} (cf.,
Figs.~\ref{fig:xdpp_liq_k0} and~\ref{fig:xtpp_liq_k0}). Thus, the
scattering from the molecules HD and HT establishes the effective
mechanism of $d\mu$ and $t\mu$ deceleration (see
Figs.~\ref{fig:xddp11_liq_k0} and~\ref{fig:xttp11_liq_k0}) and spin
deexcitation due to the presence of deuteron or triton in the target
molecules. Since the concentrations of deuterium and tritium are small,
the region of thermal energies is reached at much larger times than in
the case of pure liquid H$_2$ and D$_2$. As a result, in such H/D/T
mixtures, one can expect at large times significant effects associated
with the dynamics of the liquid.

%------------------------------------------------------------------------

\begin{acknowledgments}
  Helpful discussions with Drs. D.~L.~Demin and M.~P. Faifman are
  gratefully acknowledged.   
\end{acknowledgments}

%------------------------------------------------------------------------

%\bibliography{agliq}

%=========================================================================
\end{document}